\documentclass[aps,prl,superscriptaddress,twocolumn]{revtex4}
\usepackage{graphicx}
\usepackage{epsfig}
\usepackage{amsmath,bbm}
\usepackage{amssymb}
\usepackage{color}

\newcommand{\I}{\ensuremath{\mathrm{i}}}

\begin{document}
\title{Theory of self-induced back-action optical trapping in nanophotonic systems}
\date{\today}

\author{Lukas Neumeier}
\affiliation{ICFO - Institut de Ciencies Fotoniques, Mediterranean
Technology Park, 08860 Castelldefels (Barcelona), Spain}
\email{lukas.neumeier@icfo.es}

\author{Romain Quidant}
\affiliation{ICFO - Institut de Ciencies Fotoniques, Mediterranean
Technology Park, 08860 Castelldefels (Barcelona), Spain}
\affiliation{ICREA - Instituci\'{o} Catalana de Recerca i Estudis Avan\c{c}ats, 08010 Barcelona, Spain}

\author{Darrick E. Chang}
\affiliation{ICFO - Institut de Ciencies Fotoniques, Mediterranean
Technology Park, 08860 Castelldefels (Barcelona), Spain}
\begin{abstract}
Optical trapping is an indispensable tool in physics and the life sciences. However, there is a clear trade off between the size of a particle to be trapped, its spatial confinement, and the intensities required. This is due to the decrease in optical response of smaller particles and the diffraction limit that governs the spatial variation of optical fields. It is thus highly desirable to find techniques that surpass these bounds. Recently, a number of experiments using nanophotonic cavities have observed a qualitatively different trapping mechanism described as ``self-induced back-action trapping" (SIBA). In these systems, the particle motion couples to the resonance frequency of the cavity, which results in a strong interplay between the intra-cavity field intensity and the forces exerted. Here, we provide a theoretical description that for the first time captures the remarkable range of consequences. In particular, we show that SIBA can be exploited to yield dynamic reshaping of trap potentials, strongly sub-wavelength trap features, and significant reduction of intensities seen by the particle, which should have important implications for future trapping technologies.
\end{abstract}
\maketitle
\begin{figure}[h!,t,b]
     \centering
\includegraphics[width=8.7cm]{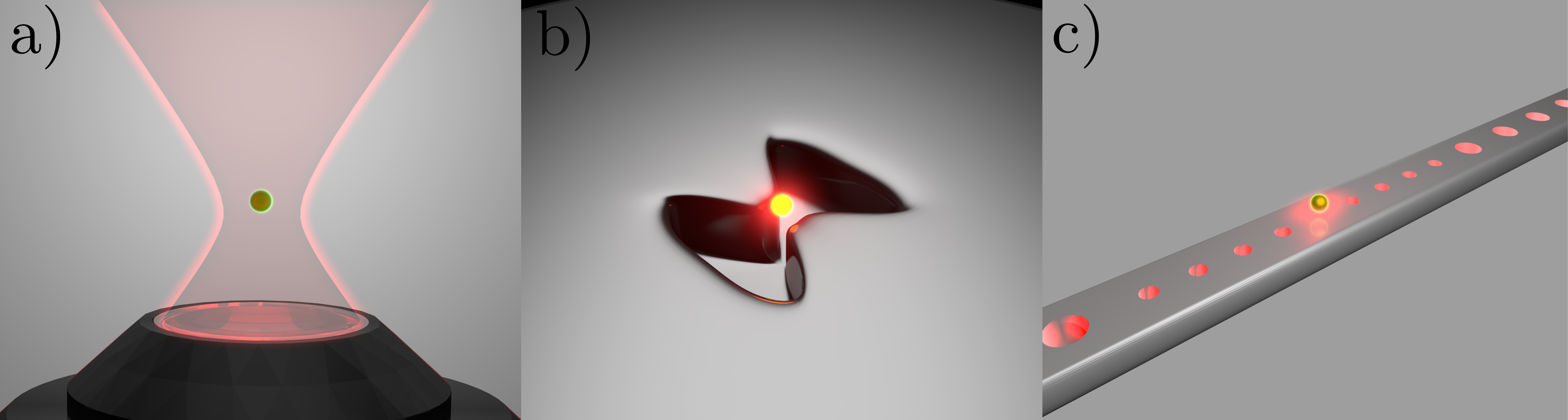}
\caption {\textbf{Schematic illustration of trap configurations.} a) A dielectric particle trapped with an optical tweezer in free space, b) a plasmonic cavity~(\textit{e.g.}, a metallic bowtie antenna), c) a photonic crystal cavity.}
\label{art}
\end{figure}

Optical trapping is one of the most important experimental tools in physics and life sciences because it enables precise control over small dielectric particles \cite{trap}.
Famous examples of its use are optical levitation and cooling of nanoscale particles \cite{levas,giseler,lev4,hyb,geraci}, trapping of bacteria \cite{bacteria}
and cells \cite{cell}, optical sorting in microfluidic channels \cite{cells}, the
manipulation and stretching of DNA \cite{dna}, and recently, even trapping of individual HIV-1 viruses \cite{virus}.
However, the difficulty of trapping a particle generally increases with decreasing size, due to the decreased optical response of the particle. This requires a commensurate increase in field intensity to maintain trap stability, and leads to associated problems such as thermal or material damage.
Another limiting factor is the diffraction limit, which constrains the length scale over which fields can vary, and thus the stiffness or possible spatial features that a trap can possess.
\\
\\
A number of experiments in recent years have migrated from trapping in free-space beams to the fields generated in nano-optical resonators \cite{nanores1,nanores2,nanores3,nanores4,SIBA,SIBA3} as illustrated in Fig.~\ref{art}. Such a paradigm can enable some technical advantages. For example, the resonator allows one to build up a higher intensity seen by the particle within the structure compared to the input, thus relaxing input power requirements. Engineering the nanophotonic structure also provides some flexibility over the field profile, and thus the trapping potential. However, it is clear that simply replacing the input field with the enhanced one does not relax any requirements from the standpoint of intensity seen by the particle. Therefore it remains an open question whether one can circumvent these seemingly fundamental trade offs between particle size and the intensities required to achieve given trap depths, frequencies, and spatial confinement. At the same time, doing so would have significant implications for optical manipulation as a tool in physics, chemistry and biology.
\\
\\
\\
In this context, a number of experiments have observed qualitatively new trapping behavior in nanophotonic cavities \cite{SIBA,SIBA3}. The key physics is that the position of the trapped particle alters the resonance frequency. This results in a ``self-induced back-action"~(SIBA) effect in which the motion dynamically affects the build up of intra-cavity intensity, and thus the optical force exerted. However, the involved trapping mechanism and its range of consequences has hardly been explored.
\\
\\
Here, we develop a general theoretical model for SIBA. Using such a model, we show how parameters can be chosen to maximize the effects of back-action, and that a single ``back-action parameter" $\eta \propto Q\frac{V}{V_m}$, proportional to the resonator quality factor and the ratio of particle to cavity mode volumes, characterizes the performance of any optimized system. In particular, the back-action parameter indicates how many line widths the particle can shift the cavity resonance frequency due to its movement.
For large $\eta$, large shifts in the cavity detuning relative to the laser frequency as the particle moves can induce strong changes in the intra-cavity intensity. Under these circumstances, and when properly optimized, such a trap yields very different trade-offs between intensities, trap depth, and confinement, which should have significant consequences for optical trapping technology. Specifically, we show that back-action can be exploited to create traps with strongly sub-wavelength spatial features, even if the cavity mode itself obeys the diffraction limit. The spatial features of the trap can also be dynamically shaped using only changes in laser frequency. Furthermore, the particle can effectively be trapped in a dynamical intensity minimum, even if it is nominally high-intensity seeking, which can strongly reduce the effects of photo-thermal damage. Finally, we discuss the possibilities for implementation in nano-plasmonic (Fig.~\ref{art}b) and photonic crystal (Fig.~\ref{art}c) systems.
\section{Model of Trapping in Nanoscale Resonators}
\begin{figure}[h!,t,b]
\centering
\includegraphics[scale=0.5]{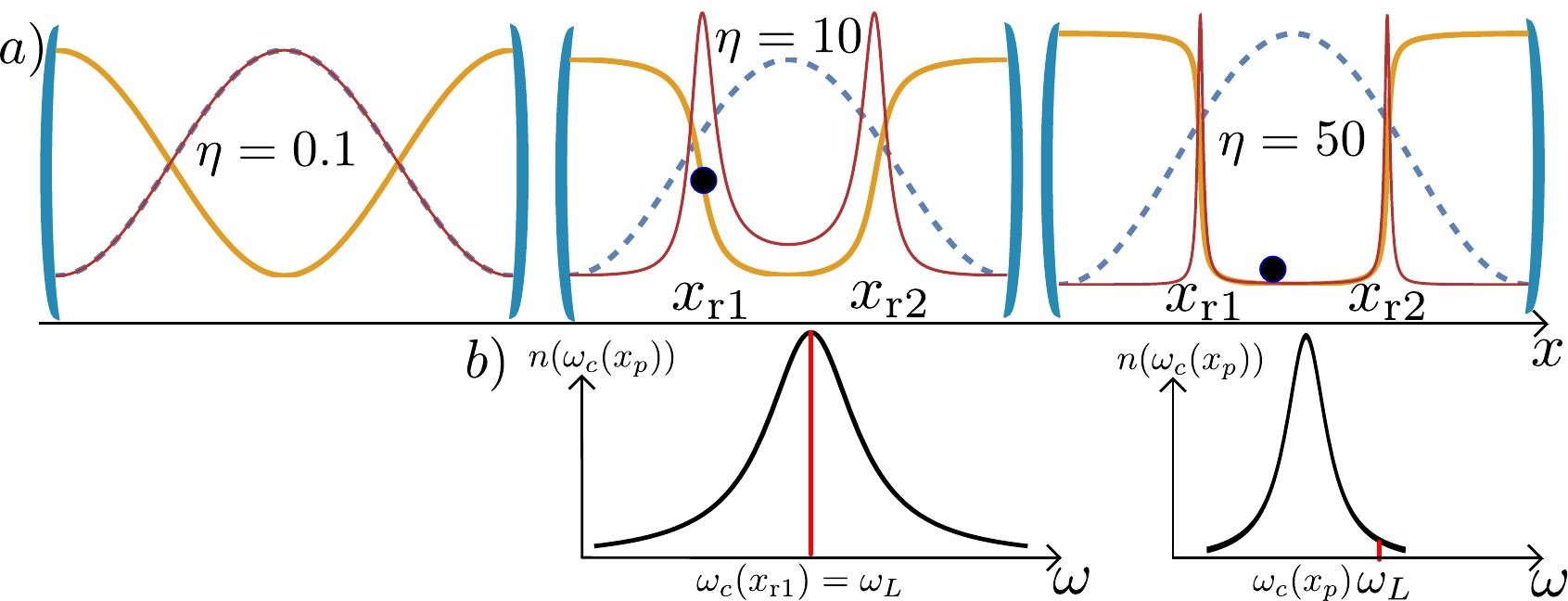}
\caption{\textbf{Back-action trapping in the fundamental mode of a Fabry-Perot cavity}, with dimensionless intensity profile $f(x)= \cos^2(kx)$ (dashed blue curve). a) In the regime of small back-action parameter ($\eta=0.1$), the intra-cavity intensity is not significantly affected by the particle motion. Thus, the local intensity $I(x_p)$ seen by the particle (red) is directly proportional to $f(x)$, while the trapping potential $U(x)\propto -f(x)$ (yellow).
For an increasing back-action parameter $\eta \gg 1$ the seen local intensity $I(x_p)$ forms sharp peaks centered around the resonant points $x_r$ and the trapping potential $U(x)$ converges to a square well potential. b) Spectra of intra-cavity photon number $n(\omega_c(x_p))$ taken at the instantaneous particle positions shown in the $\eta=10$ and $\eta=50$ cases, respectively. For $\eta=10$, we consider the case where the particle is instantaneously located at one of the resonant positions $x_{\mathrm{r1}}$, such that the laser frequency is resonant with the cavity at this moment to generate a large intra-cavity intensity. For $\eta=50$, the particle is far from the resonant positions, and the large detuning of the laser from resonance strongly suppresses intra-cavity intensity. The vertical scales of these plots are in arbitrary units.
}
\label{regimes}
\end{figure}

We first briefly review the properties and limits of trapping with free-space optical tweezers. Subsequently we will present our model of trapping in nano-resonators.
Considering a small dielectric particle whose dimensions are much smaller than the optical wavelength $d \ll \lambda$ its response is that of a point dipole with induced dipole moment $p_{\mathrm{ind}} = \alpha(\omega) E(x)$.
The well-known time averaged potential for optical trapping in free-space, such as by an optical tweezer (see Fig.~\ref{art}a) reads \cite{trap}:
\begin{equation}\label{tweezer}
U_T(x) = - \frac{1}{4} \mathrm{Re} (\alpha(\omega)) |E(x)|^2
\end{equation}
where $\alpha(\omega)$ is the frequency dependent polarizability of the particle and $|E(x)|^2 \propto I(x)$ is the peak electric field amplitude squared at the particle position $x$ and at frequency $\omega$, which is proportional to the intensity $I(x)$.
In the following, we will focus on the case where the polarizability is positive and largely frequency independent, which models well a typical dielectric particle.
In this case, the dielectric particle is trapped around points of local maximum intensity.
For sub-wavelength particles, the polarizability is proportional to particle volume, $\alpha(\omega)\propto V$. It can thus be seen that the trapping of smaller particles requires a commensurate increase in intensity to maintain a fixed trap depth. Furthermore, the spring constant around the trap minimum $x_{\mathrm{min}}$, $k_{\mathrm{spring}}=U_T''(x_{\mathrm{min}})\lesssim \frac{V \cdot I}{\lambda^2}$, in addition to being proportional to the beam intensity $I$ and particle volume $V$, is at best proportional to the inverse of the optical wavelength squared, as the diffraction limit sets this as the minimum scale over which free-space optical fields can vary.
\\
\\
We now examine the case where the particle is trapped in a nanoscale cavity. Our formalism is quite general, covering equally systems such as plasmonic and photonic crystal cavities, and trapping in vacuum or fluid environments. Qualitatively, the new feature of such a system is that the resonance frequency of the cavity depends on the particle position, enabling the particle motion to feed back on its trapping potential. We then distinguish the regimes in which this system gives rise to standard optical trapping as in Eq.~(\ref{tweezer}), versus a novel ``back-action" trapping mechanism.
\\
\\
A general model of this system is given by following Hamiltonian:
\begin{equation}\label{ham}
H = \hbar \omega_c(x_p) a^\dag a  + \hbar \sqrt{\kappa_{\mathrm{ex}}} E_0 \left( a^\dag e^{- \I \omega_L t} + a e^{\I \omega_L t} \right) + \frac{p^2}{2 m}
\end{equation}
where $\omega_c(x_p)$ is the resonance frequency of the optical cavity as a function of particle position $x_p$ and $a$ is the annihilation operator of the cavity mode. $\kappa_{\mathrm{ex}}$ denotes the decay rate of the cavity into some particular external channel (such as free-space radiation, coupling fiber, etc.), which also serves as the source of injection of photons into the cavity with number flux $E_0^2$ and frequency $\omega_L$.
The last term, $E_{\mathrm{kin}} = \frac{p^2}{2 m}$, describes the kinetic energy of a particle with momentum $p$ and mass $m$. In addition to the external coupling, we assume that the cavity has an intrinsic loss rate $\kappa_{\mathrm{in}}$, such as through material absorption or scattering losses. The total cavity linewidth is thus $\kappa=\kappa_{\mathrm{in}}+\kappa_{\mathrm{ex}}$. In principle, the particle also contributes a position-dependent loss term $\kappa(x_p)$ due to its scattering of light out of the cavity mode. While this term could be explicitly included in the analysis, this position-dependent effect is negligible under reasonable conditions as the scattering rate $\propto\frac{V^2}{\lambda^6}$ rapidly falls off for sub-wavelength sizes, as shown in the Appendix. Thus, the quality factor of the resonator is defined as $Q = \frac{\omega_c}{\kappa}$, where $\omega_c$ is the empty cavity resonance frequency.
\\
\\
The system dynamics under the Hamiltonian of Eq.~(\ref{ham}) and system losses are described by standard Heisenberg-Langevin equations \cite{elements}. As the regime of interest for trapping is far from any quantum behavior, we proceed to solve their classical expectation values.  We neglect damping of the mechanical motion and the effect of a thermal environment, which do not influence the optical force and can be added independently later on. The equations of motion then read
\begin{align}\label{exc}
& \frac{dx_p}{dt}= \frac{p}{m}  \\&
 \frac{dp}{dt} = -   n(x_p) \hbar \omega_c'(x_p)  \\&
\frac{d}{dt}\beta = \I \left(\omega_L - \omega_c(x_p) \right) \beta - \frac{\kappa}{2} \beta + \sqrt{\kappa_{\mathrm{ex}}} E_0
\end{align}
where $\beta = \langle a \rangle$ is the expectation value of the photon amplitude while $n = |\beta|^2$ is the expectation value of the photon number in the resonator.
We note that even in state-of-the-art photonic crystal cavities, the achievable quality factor $Q=\frac{\omega_c}{\kappa} \approx 10^6$ results in decay times of $\kappa^{-1}\sim 1$ ns that are significantly shorter than the timescales of motion \cite{crystal}. Thus, this motivates an approximation $\frac{d\beta}{dt}\approx 0$ where the cavity is able to instantaneously respond to the particle motion.
\\
\\
Before solving equations (3)-(5), we want to examine how strongly the particle affects the resonance frequency.
To quantify this, we compare the frequency shift $\delta \omega_c(x_p)= \omega_c(x_p)-\omega_c$ with half of the line width $\frac{\kappa}{2}$ of the resonator, where $\omega_c$ is the resonance frequency of the empty cavity.
Within lowest order perturbation theory, where the particle induces a frequency shift much smaller than the bare cavity frequency, it can be shown that~(Appendix)
\begin{equation}\label{cc}
\frac{2 \delta \omega_c(x_p)}{\kappa} =  - \eta \cdot f(x).
\end{equation}
Here, we have defined the dimensionless back-action parameter $\eta = \frac{ \alpha(\omega)}{\epsilon_0 V_m}  Q$, where $V_m$ is the cavity mode volume. $f(x)$ is the dimensionless spatial intensity profile of the empty cavity, normalized to be 1 at the intensity maximum. Thus, as $0\leq f(x) \leq 1$, the back-action parameter $\eta$ characterizes how many linewidths (half-width half-maxima) the particle can shift the resonance frequency of the cavity moving from the minimum  $f(x)=0$ to the maximum of the mode profile. For sub-wavelength dielectric particles the polarizability
$\alpha(\omega)\propto \epsilon_0 V$ is proportional to the particle volume, with the pre-factor depending on the particle refractive index and shape \cite{scat}. Thus, achieving a large back-action parameter requires a sufficient combination of large cavity quality factor and ratio of particle to cavity mode volume, $\eta \propto \frac{V}{V_m} \cdot Q$. When the particle size is larger than $kr\gtrsim Q^{-\frac{1}{6}}$~(with $k=2\pi/\lambda$), the effect on the quality factor due to light scattering by the particle cannot be neglected anymore, and this case is considered further in the SI.
\\
\\
The expectation value of the intra-cavity photon number $n(x_p)=|\beta|^2$ reads:
\begin{equation}\label{number}
n(x_p) = \frac{4 E_0^2 \kappa_{\mathrm{ex}}}{\kappa^2} \frac{1}{1 + \left( \eta f(x_p) + \tilde{\Delta} \right)^2}
\end{equation}
where we have defined the dimensionless detuning between the laser and empty cavity frequencies, $\tilde{\Delta} = \frac{2(\omega_L-\omega_c)}{\kappa}$. From Eq.~(\ref{number}), one sees that there are certain positions of the particle $x_r$ that cause the driving laser to become resonant with the (frequency-shifted) cavity, $\tilde{\Delta}+\eta f(x_r)=0$, and where the intra-cavity photon number is maximized.
We call these positions  the resonant positions $x_r$, which can be chosen by adjusting the laser frequency $\omega_L$.
Note that in arbitrary dimensions the resonant positions $\vec{x}_r$ are contour points/lines/surfaces in 1D/2D/3D and follow the symmetry of the mode profile, see Fig.~\ref{regimes}.
Inserting Eq.~(\ref{number}) into Eq.~(4) and integrating the negative force with respect to $x_p$
yields the general potential for trapping in resonators:
\begin{equation}\label{pot}
 U(x) = -2 \hbar E_0^2 \frac{\kappa_{\mathrm{ex}}}{\kappa} \arctan\left[\eta f(x)+ \tilde{\Delta}\right].
\end{equation}
We will proceed by looking at different regimes of this potential:
First we consider the regime where the particle induces a shift on the cavity resonance frequency that is negligible compared to its linewidth, which corresponds to $\eta \ll 1$ from our definition in Eq.~(\ref{cc}). Then, the movement of the particle does not significantly change the intra-cavity intensity, which recovers the optical tweezer regime. In particular,
expanding Eq.~(\ref{pot}) for small $\eta$, one finds that $U_T(x)= -  2 \hbar \frac{\kappa_{\mathrm{ex}}}{\kappa}  E_0^2  \frac{\eta}{1+\tilde{\Delta}^2} f(x)$. Using the definition of $\eta= \frac{\alpha(\omega)}{\epsilon_0 V_m} Q$ and identifying $|E(x)|^2= 8  \frac{\hbar}{\epsilon_0 V_m} \frac{\kappa_{\mathrm{ex}}}{\kappa} Q E_0^2 \frac{1}{1+\tilde{\Delta}^2} f(x)$ as the time averaged intra-cavity field amplitude, we see that $U_T(x)$ reduces to the optical dipole potential in Eq.~(\ref{tweezer}). In this regime, the potential depth increases linearly with $Q$~(i.e., with $\eta$), reflecting the effect of a built-up intra-cavity intensity.
The different regimes are illustrated in Fig.~\ref{regimes} where we choose the first harmonic of a Fabry-Perot cavity as a mode profile.
\section{Trapping with back-action}
\begin{figure}[h!,t,b]
     \centering
\includegraphics[width=8.7cm]{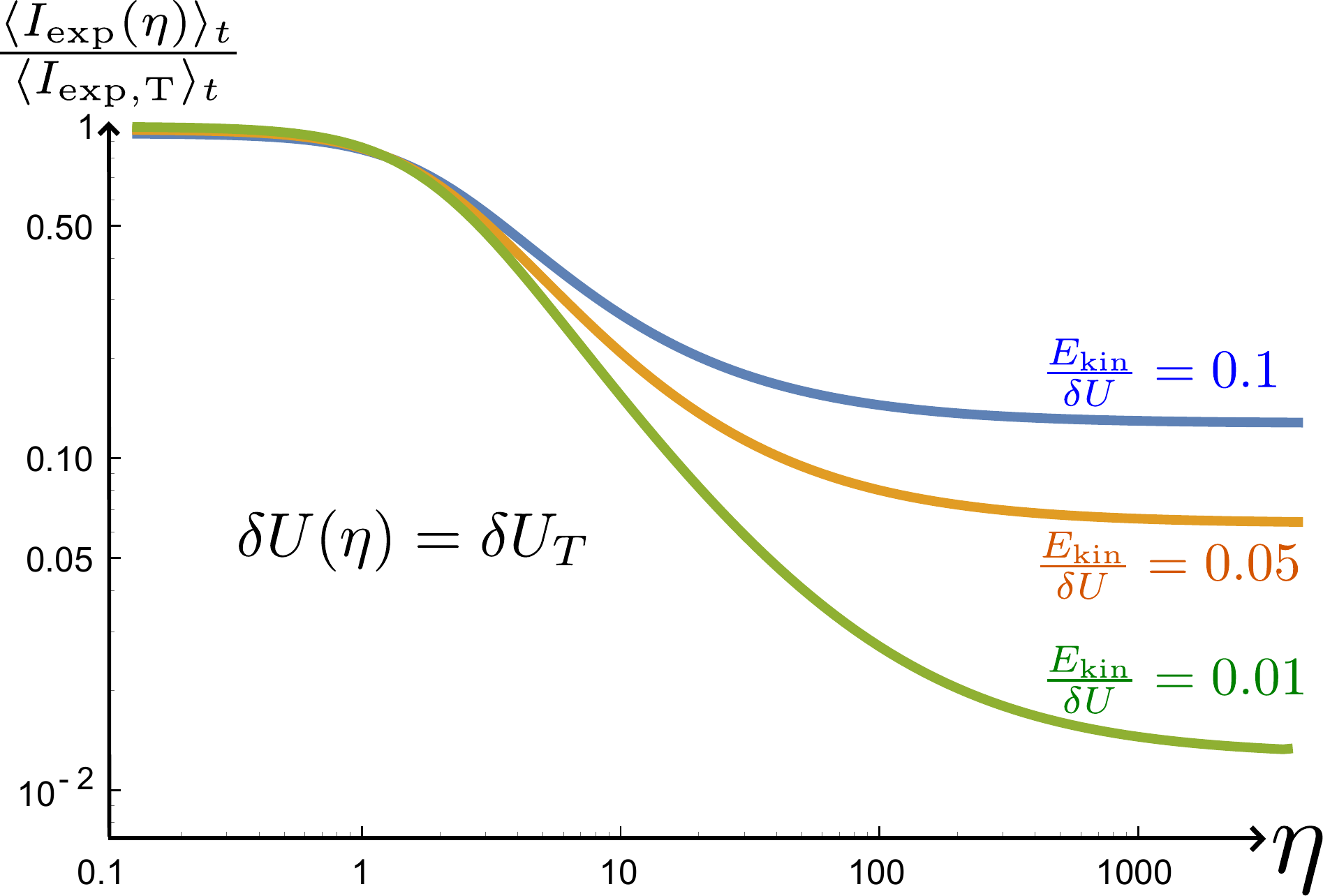}
\caption {\textbf{Time-averaged experienced intensity of a trapped particle.} We plot the time-averaged experienced intensity as a function of back-action parameter $\langle I_{\mathrm{exp}}(\eta) \rangle_t$, normalized with the value in the optical tweezer regime $\eta \ll 1$. The two cases are set to have equal trap depth. The plot is numerically calculated for the case of trapping in the fundamental mode of a Fabry-Perot cavity $f(x) = \cos^2(kx)$ with resonant positions $k x_r= \frac{\pi}{4}$. The back-action regime can enable much lower average local intensities than in the optical tweezer regime.}
\label{compare}
\end{figure}

We now investigate the very different trap properties that emerge in the regime $\eta\gg 1$.
\\
\\
An increase in the quality factor initially produces an increased trap depth for values $Q \lesssim \pi \frac{V_m}{V}$~(at which point $\eta\sim 1$). For larger values, however, $\eta \gg 1$ and the arctan in Eq.~\ref{pot} saturates between the values of $\pm \frac{\pi}{2}$, yielding a trap depth of $\delta U= 2 \pi \hbar E_0^2 \frac{\kappa_{\mathrm{ex}}}{\kappa}$.
Significantly for $\eta \gg 1$, the depth no longer depends on $Q$ nor the particle properties, and is only dependent upon the input intensity.
The origin of this saturation can be understood by first considering Fig.~\ref{regimes}, which shows that the intra-cavity intensity as a function of particle position forms
sharp peaks around the resonant positions $x_r$ for $\eta \gg 1$. From Eq.~(7) it follows that their width is in good approximation $\approx \frac{2}{\eta f'(x_r)}$ and it is only within this narrow spatial region (scaling like $\eta^{-1}\propto Q^{-1}$) that the cavity exerts significant forces on the particle. At the same time, the peak intra-cavity photon number at $x_p=x_r$ (and thus the peak force) grows linearly with $Q$. Thus, the maximum work that the cavity can do to keep the particle in the trap, as a product of force and distance, becomes independent of $Q$ in the high-back-action limit. \\
\\
Note that the trapping potential turns into an approximate square well if the distance between the intra-cavity intensity peaks is larger than their width $d=|x_{\mathrm{r2}}-x_{\mathrm{r1}}| \gg \frac{2}{\eta f'(x_r)}$. The wells are (symmetrically) centered around the mode profile maximum $x_0$, see Fig.~\ref{regimes}.
Remarkably, the resonant positions $x_r$ can be changed with laser frequency, which provides a convenient mechanism for dynamic trap shaping in contrast with conventional optical tweezers.
\\
\\
Another interesting property of the trap in the high back-action regime is that around the minimum $x_0$ of the potential, the intra-cavity photon number is strongly suppressed due to the large detuning from resonance. Thus, the particle is effectively trapped in a dynamical intensity minimum, despite the fact that it has positive polarizability and is thus nominally high-intensity seeking. This would have tremendous consequences in the reduction of thermal damage due to optical absorption by the particle. Motivated by this observation, we seek to quantify how much the time-averaged intensity seen by the particle can be reduced.
\\
\\
We define the time-averaged experienced intensity $ \langle I_{\mathrm{exp}} \rangle_t$  as the local intensity experienced by the particle at its position, averaged over one motional period $T$. It is thus given by
\begin{equation}\label{ing}
 \langle I_{\mathrm{exp}} \rangle_t =   \frac{c \hbar  \omega_L }{2 V_m T} \int_0^T  n(x_p(t)) f(x_p(t)) dt
\end{equation}
where $x_p(t)$ is a solution to the differential Eq.~(4) together  with Eq.~(7).
In order to proceed further, we consider a simple case of the fundamental mode of a 1D Fabry-Perot cavity, $f(x) = \cos^2(kx)$ with $k=\frac{\pi}{L}$, where $L=\lambda/2$ is the cavity length. Although we have switched to a specific model to illustrate the back-action mechanism, we believe the overall conclusions are generally valid.
A finite temperature of the environment can be taken into account by averaging the results for different maximal kinetic energies $E_{\mathrm{kin}}$ (kinetic energy of the particle in the trap minimum) according to a Boltzmann distribution.
\\
\\
We have evaluated Eq.~(\ref{ing}) by numerically solving the equations of motion (3)-(5). In Fig.~\ref{compare}, we plot the time-averaged experienced intensity $ \langle I_{\mathrm{exp}}(\eta)\rangle_t$ normalized by the value in the optical tweezer regime $\langle I_{\mathrm{exp},T}\rangle_t$, as a function of back-action parameter $\eta$. As seen before, the optical tweezer regime is reached by taking $\eta \ll 1$.
To make a fair comparison, we enforce that the trap depths in the two cases are equal, $\delta U(\eta) = \delta U_T$.
For a fixed $x_r$, the figure shows a significant reduction in time-averaged intensity for high back-action parameter, which also depends on the ratio of kinetic energy $E_{\mathrm{kin}}$ to trap depth $\delta U$. In the high back-action regime, it is possible to derive an analytic expression (Appendix):
\begin{equation}\label{an}
 \lim_{\eta \rightarrow \infty }  \langle I_{\mathrm{exp}}(\eta) \rangle_t = \frac{2c \epsilon_0}{\alpha(\omega)}\frac{f(x_r)}{|f'(x_r)|} \frac{E_{\mathrm{kin}}}{x_r}
\end{equation}
A new feature of the back-action trap is the gradual decoupling between trap depth and the spatial region $\delta x = |x_{\mathrm{t2}}-x_{\mathrm{t1}}|$ ($x_{\mathrm{t1}}$ and $x_{\mathrm{t2}}$ are the classical turning points) to which the particle is confined. For large enough $\eta$ they decouple completely since the classical turning points converge to the resonant positions (i.e., the edges of the square well) and thus $\delta x \rightarrow  d$. In this regime, confinement only depends on laser frequency, whereas trap depth only depends on laser power. This independent control again highlights the ability to dynamically reshape the trap. In contrast, in the optical tweezer regime, the trap depth, kinetic energy and confinement are inevitably connected.
\\
\\
Instead of comparing the experienced intensity at fixed trap depth, we can also investigate the trade-off between intensity and confinement $\delta x = d$ in the large back-action limit.
The locations of the trapping wells are always centered around the mode profile maximum $x_0=0$.
For small $x_r$, an asymptotic expansion yields $\frac{f(x_r)}{|f'(x_r)|} \approx \frac{1}{2 k^2 x_r}$.
Thus, for high back-action and strong confinement, we obtain $ \langle I_{\mathrm{exp}} \rangle_t  \approx \frac{4 c \epsilon_0}{\alpha(\omega)}  \frac{1}{(k \delta x)^2} E_{\mathrm{kin}}$. Interestingly, expanding Eq.~(1) for the optical tweezer around the bottom of a standing wave potential also produces  $\langle I_{\mathrm{exp,T}} \rangle_t \approx I(x_0)  \approx  \frac{4 c \epsilon_0}{\alpha(\omega)}  \frac{1}{(k \delta x)^2} E_{\mathrm{kin}}$, which seems to indicate that no improvement is gained in intensity vs. confinement with back-action.
\\
\\
Looking at Eq.~(10), in the strong back-action regime, one of the factors of $\frac{1}{\delta x}$ originates simply from the time $T\propto\delta x$ that the particle takes to travel between the walls of the square well. This part of the scaling seems fundamental and cannot be improved within this model.
On the other hand, the second factor of $\frac{1}{k \delta x}$ clearly originates from the vanishing of back-action effects around the maximum of the mode profile, as the frequency shift becomes insensitive to first-order changes in the particle displacement, $f'(x_0)=0$.  We show that this factor is not fundamental, and can be eliminated by properly driving a second optical mode of the system.
 \section{Two Mode Back-action}
\begin{figure}[h!,t,b]
\centering
\includegraphics[width=8.7cm]{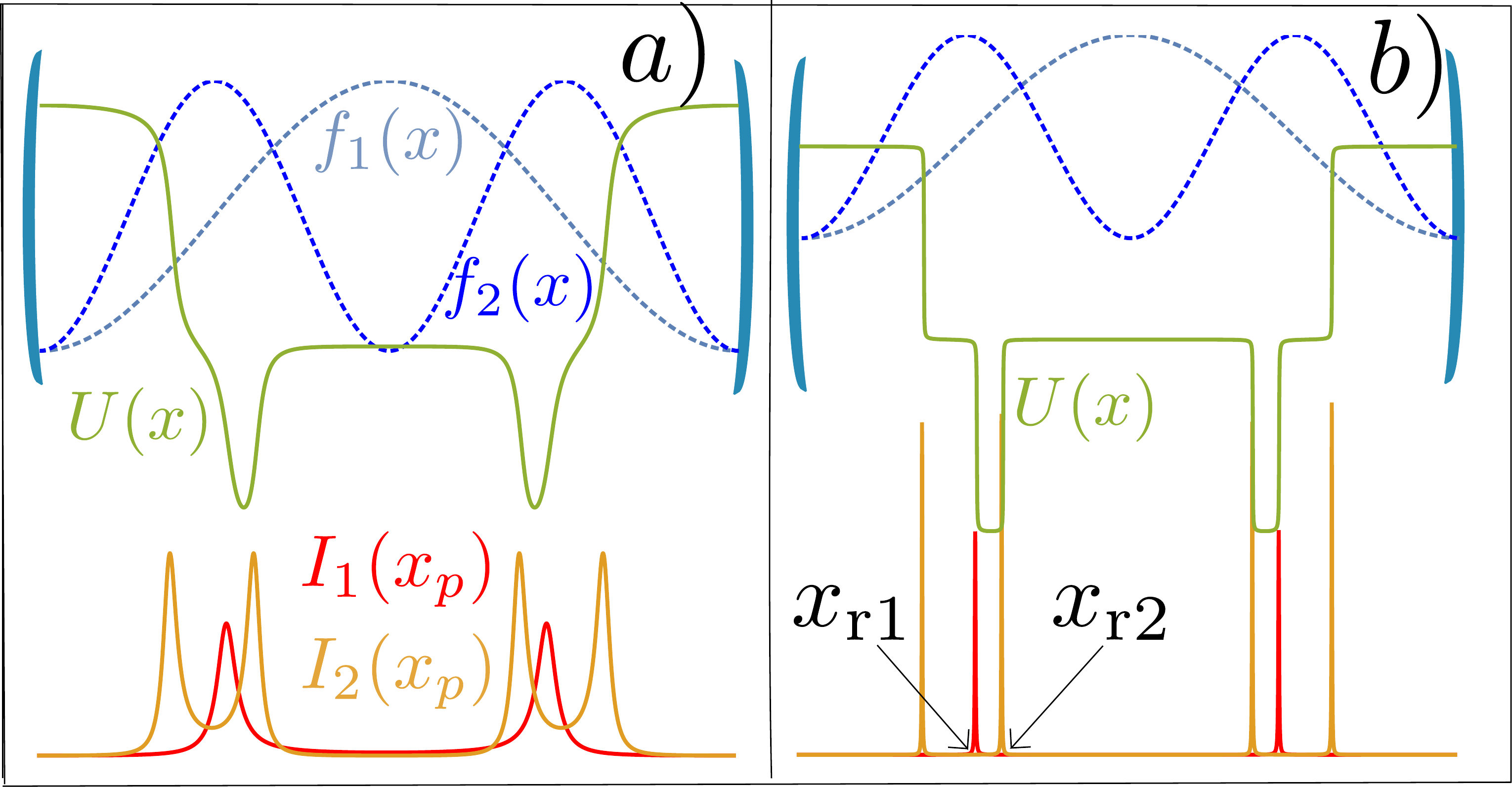}
\caption{\textbf{SIBA with two optical modes}, illustrated here for the first two modes of a Fabry-Perot cavity. Top: the mode profiles are given by $f_1(x)=\cos^2(kx)$, $f_2(x) = \sin^2(2kx)$. Green: the resulting optical trapping potential $U(x)$.
Bottom: intra-cavity intensities $I(x)$ as a function of particle position. $a)$ In the harmonic back-action regime, the distance between the resonant points is comparable to the width of the intensity peaks, $ k d \sim \frac{2}{\eta}$. $b)$ In the high back-action regime, the distance significantly exceeds the width, $ k d  \gg \frac{2}{\eta}$.}
\label{twomodepot}
\end{figure}

\begin{figure*}[t]
\centering
\includegraphics[width=18cm]{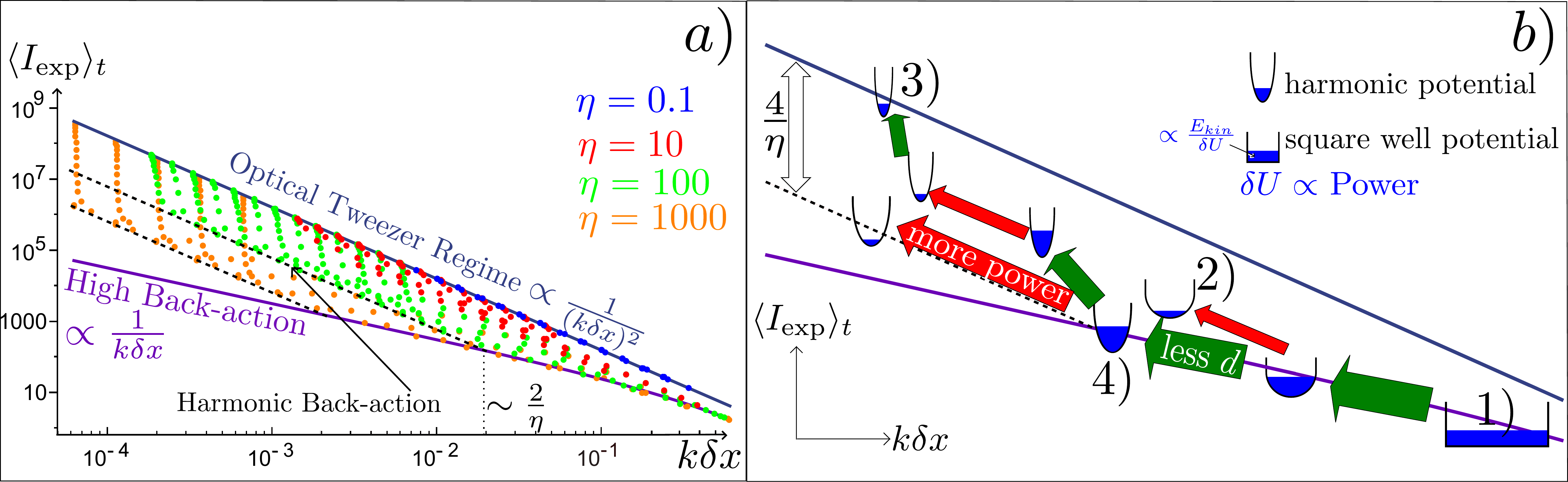}
\caption{\textbf{Time-averaged experienced intensity vs. confinement for two optical modes.} a)
Time-averaged experienced intensity in units of $\frac{c E_\mathrm{kin}}{\alpha(\omega)}$ as a function of confinement $k \delta x = k|x_{\mathrm{t2}}-x_{\mathrm{t1}}|$. The individual points originate from different combinations of back-action parameter, laser power and detunings. The solid lines indicate the scalings in the optical tweezer regime and high back-action regime, where $kd=k|x_{\mathrm{r2}} - x_{\mathrm{r1}}|\gg \frac{2}{\eta}$. The dashed line shows the optimized harmonic back-action regime, where $kd\sim \frac{2}{\eta}$.
b) Illustration of figure \ref{mus}a for a fixed $\eta$ and $E_{\mathrm{kin}}$, and a schematic of the protocol to saturate the scaling bounds.
The green arrows denote a decrease in distance between the resonant positions $d= |x_{\mathrm{r1}}-x_{\mathrm{r2}}|$, while the red arrows denote an increase in laser power. The ratio of filled blue area to whole area of the parabolae/square wells indicates the ratio of $\frac{E_{\mathrm{kin}}}{\delta U}(\propto \frac{1}{\textrm{Power}})$. 1) high back-action regime: $d$ and $\delta U$ (laser power) decouple, the potential forms a square well and
$\delta x \rightarrow d$. 2) a significant increase in laser power prevents the particle from exploring the full square well potential. 3) decreasing $kd<2/\eta$ suppresses back-action as the particle motion no longer shifts the cavity mode frequencies. 4) beginning of the optimized harmonic back-action regime. Maximum confinement at this point is achieved when $\frac{E_{\mathrm{kin}}}{\delta U} \approx \frac{1}{2}$. Maintaining the harmonic back-action scaling~(dashed line) is achieved by increasing laser power.
\label{mus}}
\end{figure*}

In this section we show how the scaling between experienced intensity and confinement can be improved to $\langle I_{\mathrm{exp}} \rangle_t \propto \frac{1}{k \delta x}$ by using two different cavity modes for trapping. In order to obtain concrete results, we consider the simple geometry where the two modes consist of the first and second harmonics of a Fabry-Perot~(see Fig.~\ref{twomodepot}), although we believe that the conclusions hold quite generally. We assume that each mode can be driven with its own laser, with amplitude $E_{\mathrm{0}i}$ and frequency $\omega_{\mathrm{L}i}$. As the equation for the intra-cavity fields $\beta_i$~(generalized from Eq.~(5)) of each mode are decoupled from one another, they can be separately integrated as in the single-mode case. Thus, the total potential $U_{\mathrm{tot}}(x)=\sum_{i=1,2}U_{i}(x)$ is the incoherent sum of the potentials in Eq.~(9) for each mode. To understand the relevant physics, it is sufficient to assume that the mode driving amplitudes $E_{\mathrm{0i}}$, decay rates $\kappa_{\mathrm{ex}},\kappa_{\mathrm{in}}$, and back-action parameters are identical, although the concepts can be easily generalized.
 \\
 \\
The interesting regime will be when the resonant positions of each mode are tuned by their respective driving laser frequencies such that each mode is responsible for providing one trapping wall. This is illustrated in Fig.~\ref{twomodepot}b, where the left and right walls $x_{\mathrm{r1}}$ and $x_{\mathrm{r2}}$ originate from the first and second cavity modes, respectively. Significantly, the well can be located far from the nodes/antinodes $f'_i(x)=0$ where the effects of back-action would vanish for either mode. In the following we will distinguish three different regimes concerning the ratio between the distance $d=|x_{\mathrm{r1}}- x_{\mathrm{r2}}|$ and the width $\sim \frac{2}{k\eta}$ of these intensity peaks illustrated in Fig.~\ref{twomodepot}.
\\
\\
We start by examining the high back-action regime, when the distance of the intensity peaks is much larger than their width, $ kd = k |x_{\mathrm{r1}}-x_{\mathrm{r2}}| \gg \frac{2}{\eta}$, such that we encounter an almost perfect square-well potential as shown in Fig.~4b. It is straightforward to generalize the high back-action limit of Eq.~(\ref{an}) in the single mode case.
As the particle is trapped far from points where back-action effects vanish ($f_i'(x)=0$), we recover the improved scaling between experienced intensity and confinement, $ \langle I_{\mathrm{exp}} \rangle_t \propto \frac{1}{k \delta x}$ as already anticipated.
\\
\\
In Fig.~\ref{mus}, we have illustrated the results of experienced intensity vs.~confinement from full numerical simulations of Eqs.~(3)-(5) (generalized to two modes). Here, the different points for a fixed back-action parameter $\eta$ are obtained by variation of the input powers and resonant positions $x_r$ (via the laser frequencies). Tuning the resonant positions to reduce $d= |x_{\mathrm{r1}}-x_{\mathrm{r2}}|$ indeed enables one to saturate the scaling of $\langle I_{\mathrm{exp}}\rangle_t \propto \frac{1}{k \delta x}$ as long as $k\delta x \gtrsim \frac{2}{\eta}$, as illustrated in Fig.~\ref{mus}b.
\\
\\
For $k\delta x\lesssim \frac{2}{\eta}$, the optimal scaling seen in the numerics goes like $\langle I_{\mathrm{exp,hb}}\rangle_t \propto \frac{1}{\eta(k\delta x)^2}$. The scaling with $\delta x^{-2}$ resembles the optical tweezer case, but the intensity is suppressed by a factor of $\eta$. We call this the ``harmonic back-action regime''~(see Fig.~\ref{twomodepot}a). To understand this case, we first note that the particle moves by a small enough amount around the trap minimum that the forces from each mode can be linearized around small displacements to yield a harmonic trap. Furthermore, for small displacements, the total time averaged experienced intensity  $\langle I_{\mathrm{exp}} \rangle_t = \sum_i \langle I_{\mathrm{exp},i} \rangle_t \approx \sum_i I_i(x_0)$ is just the sum of the intensities of the respective mode at the trap minimum $x_p = x_0$. The associated spring constant is:
 \begin{equation}\label{optical}
 k_{\mathrm{opt}} = -F'(x_0) = \sum_i n_i'(x_0) \omega'_{c,i}(x_0) + n_i(x_0) \omega_{c,i}''(x_0)
\end{equation}
where the sum goes over all trapping modes.
The first term $n_i'(x_0)$ is a new contribution to the optical spring constant $k_{\mathrm{opt}}$ originating from the change in photon number with particle position around the trap minimum.
Intuitively, this back-action contribution to the spring constant is maximized by ensuring the photon number of each mode maximally changes around $x_0$. This is roughly optimized by setting $kd \sim  \frac{2}{\eta}$, such that $x_0$ corresponds to sitting half a cavity linewidth away from the resonant position $x_r$. Such an optimization yields (Appendix):
\begin{equation}\label{kop}
 k_{\mathrm{opt},i} =  \frac{\alpha(\omega)}{ c \epsilon_0} \langle I_{\mathrm{exp},i} \rangle_t  \frac{1}{f_i(x_0)} \left[ \eta_i f_i'(x_{\mathrm{r}i})^2- f_i''(x_0)\right].
\end{equation}
The first term in the brackets originates from the change in photon number with particle position, whereas the second term reduces to the optical tweezer spring constant given by Eq.~(\ref{tweezer}): $k_{T}=U''(x_0)$.
Since $f''(x_0) \sim f'(x_0)^2 \sim k^2$, it can be seen that the back-action contribution is a factor of $\eta$ larger. We can equivalently interpret this contribution as arising from an effective reduced wavelength $\lambda_{\mathrm{eff}} \sim \frac{\lambda}{\sqrt{\eta}}$, which enables the generation of trap features far below the diffraction limit. We emphasize that this effect originates from the rapid change in intra-cavity photon number with particle displacement rather than a change in the spatial mode itself (see Eq.~(\ref{optical})), and thus there is no breakdown of the dipole approximation in which all of these expressions are derived.
It should be noted that an analogous ``optical spring" effect has been reported in optomechanical systems \cite{op1,op2}, where an optical cavity can exert large restoring forces for small displacements of a mechanical system. Here, the stiffness of the mechanical mode itself plays the role of our second optical mode, and serves to keep the equilibrium position at a point of non-vanishing back-action ($f'(x_0)\neq 0$) \cite{op1}.
\\
\\
Exploiting the notion of a reduced wavelength, in the harmonic back-action regime one can immediately conclude that the scaling for average experienced intensity improves from  $\langle I_{\mathrm{exp},\mathrm{T}}\rangle_t \propto \frac{1}{(k\delta x)^2} $ for an optical tweezer to  $\langle I_{\mathrm{exp},\mathrm{hb}}\rangle_t \propto \frac{1}{\eta ( k \delta x)^2}$. A more detailed optimization of the system shown in Fig. 3 and explained in the SI reveals that:
\begin{equation}
  \frac{ \langle I_{\mathrm{exp,hb}}\rangle_t}{ \langle I_{\mathrm{exp,T}} \rangle_t} = \frac{4}{\eta}
 \end{equation}
for equal confinement and kinetic energy. We want to emphasize that to reach this optimal scaling, one should fix $kd = k|x_{\mathrm{r1}}-x_{\mathrm{r2}}| \sim \frac{2}{\eta}$. In this way, one stays on the dashed line scaling shown in Fig.~\ref{mus} and one achieves smaller confinement by turning up laser intensity while still maintaining the full back-action advantage. In contrast, Fig.~\ref{mus} also shows that by decreasing the distance between the resonant positions, $kd  \ll \frac{2}{\eta}$, the scaling deviates back towards the optical tweezer limit and the benefits of back-action vanish.
\section{Conclusion}
 There have already been two types of systems, plasmonic cavities \cite{romain} and photonic crystal cavities \cite{crystal}, where SIBA has already been observed, and we now discuss the potential 
 s of merit associated with each. As the plasmon resonances associated with small metallic systems do not obey a diffraction limit, they are able to achieve strongly sub-wavelength mode volumes. On the other hand, realistic quality factors are limited to $Q\lesssim 10-10^2$. At the same time, an upper bound on the validity of our calculation is that the particle size $V\lesssim V_m$ does not exceed the mode volume, and thus we anticipate maximum possible values of $\eta\sim 10-10^2$ for such systems. In photonic crystal cavities, the mode volume is limited by the diffraction limit to $V_m\gtrsim (\frac{\lambda}{2})^3$, while extremely high quality factors of $Q \sim 10^6$ are possible \cite{crystal}. This yields $\eta \sim 10, 100, 400$ for a dielectric sphere with radius $r \sim 6.5$nm,$15$nm, $28$nm~(Appendix). There has been significant activity in recent years to develop design principles in order to tailor the spatial modes of plasmonic \cite{romain} and photonic crystal structures \cite{shape} for trapping. Combined with the potentially large back-action parameters achievable, we anticipate that our work will open up significant new opportunities for optical trapping. Finally, it would also be interesting to explore the use of large back-action parameter in other functionalities, such as particle detection and feedback cooling.

\section{Acknowledgements}

The authors acknowledge stimulating discussions with M.D. Lukin, M.L. Juan, and H.J. Kimble, and the help of J. Berthelot in producing the artwork. This work was supported by Fundacio Privada Cellex Barcelona, Severo Ochoa PhD Fellowship, the MINECO Ramon y Cajal Program, the Marie Curie Career Integration Grant ATOMNANO, and ERC Starting Grants FoQAL and Plasmolight.

\end{document}


\title{Appendix to ``Theory of self-induced back-action optical trapping in nanophotonic systems''}
\date{\today}

\author{Lukas Neumeier}
\affiliation{ICFO - Institut de Ciencies Fotoniques, Mediterranean
Technology Park, 08860 Castelldefels (Barcelona), Spain}
\email{lukas.neumeier@icfo.es}

\author{Romain Quidant}
\affiliation{ICFO - Institut de Ciencies Fotoniques, Mediterranean
Technology Park, 08860 Castelldefels (Barcelona), Spain}
\affiliation{ICREA - Instituci\'{o} Catalana de Recerca i Estudis Avan\c{c}ats, 08010 Barcelona, Spain}

\author{Darrick E. Chang}
\affiliation{ICFO - Institut de Ciencies Fotoniques, Mediterranean
Technology Park, 08860 Castelldefels (Barcelona), Spain}
\begin{abstract}
Here we provide additional supporting calculations that were not included in the main text.
\end{abstract}

\maketitle

\section{1 Frequency Shift}
For small frequency shifts $\delta\omega_c(x_p)=\omega_c(x_p)- \omega_c$ compared to the unaltered resonance frequency $\omega_c$ of the cavity, we can obtain $\delta\omega_c(x_p)$ from electromagnetic perturbation theory~\cite{molding}:

\begin{myequation}
\delta\omega_c(x_p) =  -\frac{\omega_c}{2} \frac{\int d^3r  \vec{P}(r) \cdot \vec{E}(r)}{\int d^3r \frac{d}{d\omega} \left(\epsilon(\omega,r) \omega \right) \left| E(r) \right|^2}
\end{myequation}
where $\vec{E}(r)$ is the electric field of the empty resonator, $\epsilon(\omega,r)$ is the dielectric function of the empty resonator, and $\vec{P}(r)$ is the additional polarization due to the presence of the particle.
If we take the particle to be small compared to the wavelength of the laser, the electric field across the particle is approximately constant and its response is equivalent to a point dipole with polarizability $\alpha(\omega)$. As an example, for a dielectric sphere of volume $V$ and refractive index $n$ in vacuum, the polarizability can be exactly calculated, $\alpha(\omega)=3\epsilon_{0}V\frac{n^2-1}{n^2+2}$. For a given polarizability, one finds
\begin{myequation}\label{cc}
 \delta\omega_c(x_p) = -\frac{ \omega_c \alpha(\omega)}{2 \epsilon_0 V_m} f(x)
\end{myequation}
where
$f(x) = \frac{ \frac{d}{d\omega} \left( \epsilon(\omega,x) \omega \right) \left| E(x)\right|^2}{\max_x{  \frac{d}{d\omega} \left( \epsilon(\omega,x) \omega \right)} \left| E(x)\right|^2 }$. $f(x)$ describes the dimensionless spatial intensity profile of the empty cavity, normalized to be 1 at the intensity maximum.
$V_m$ is the mode volume of the empty resonator and is defined as follows:
\begin{myequation}
 V_m = \frac{\int d^3x \frac{d}{d\omega} \left(\epsilon(\omega,x) \omega \right) \left| E(x) \right|^2}{\max_x{  \frac{d}{d\omega} \left( \epsilon(\omega,x) \omega \right)} \left| E(x)\right|^2}
\end{myequation}
\section{2 Scattering Rate of the trapped particle}
\begin{figure}[h!,t,b]
\centering
\includegraphics[scale=0.75]{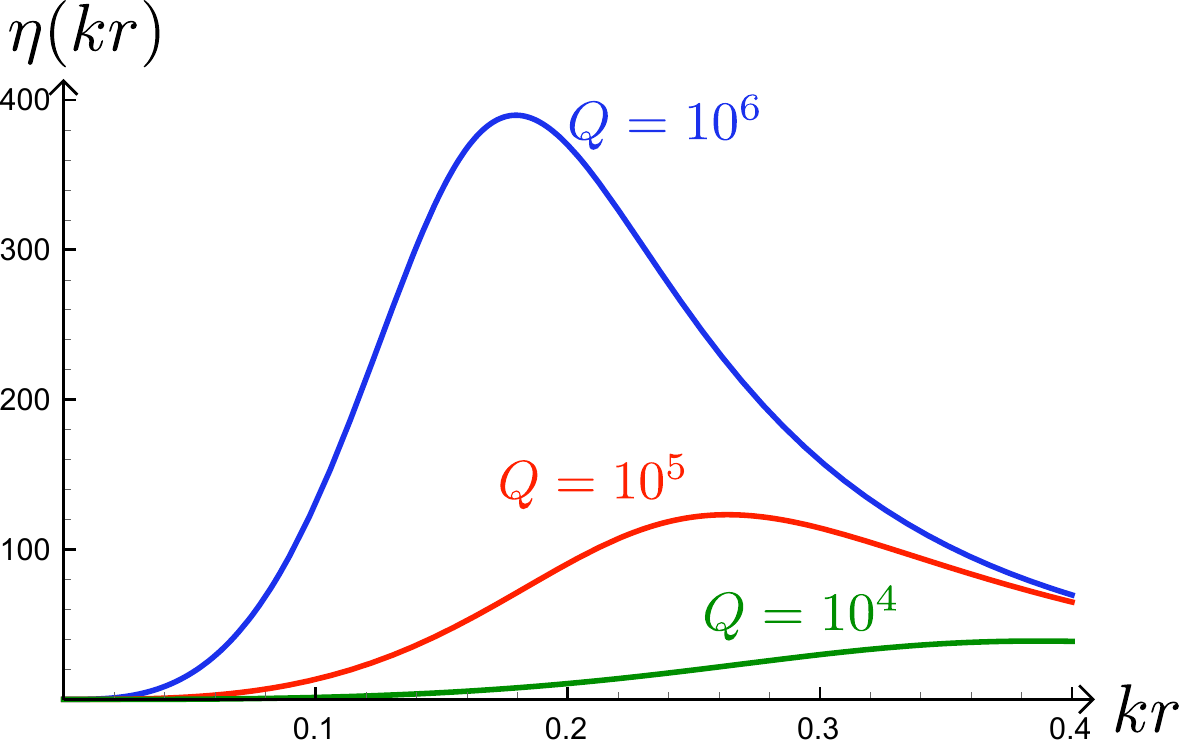}
\caption{Plot of the back-action parameter as a function of particle size after considering particle induced scattering losses as described by Eq.~(\ref{op}). Here we take $\nu = 1$ and empty-resonator quality factors of $Q=10^4,10^5,10^6$.}
\label{max1}
\end{figure}

\begin{figure}[h!,t,b]
\centering
\includegraphics[scale=0.75]{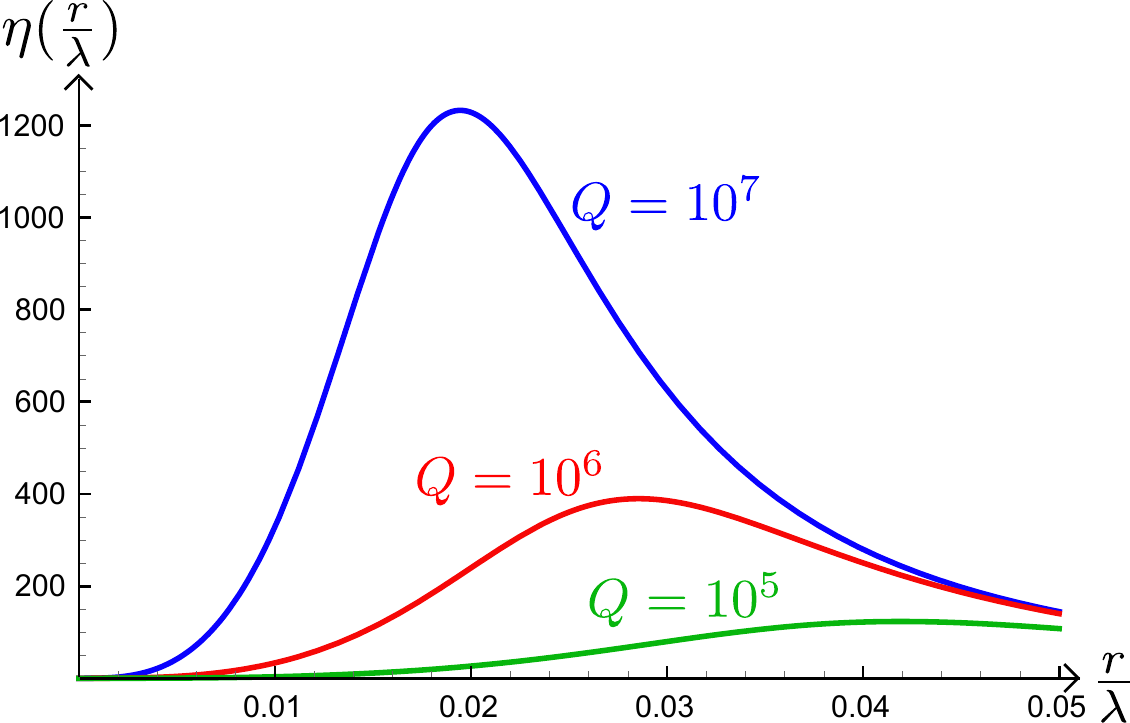}
\caption{Plot of the back-action parameter as a function of particle size after considering particle induced scattering losses as described by Eq.~(\ref{op}). Here we take $\nu = 1$ and empty-resonator quality factors of $Q=10^5,10^6,10^7$.}
\label{max2}
\end{figure}

Here we consider the scattering of light by the trapped object itself, which
decreases the cavity quality factor by contributing to its loss rate $\kappa = \kappa_{\mathrm{ex}} + \kappa_{\mathrm{int}} + \kappa_{\mathrm{scat}}(x_p)$.
The scattering rate for sub-wavelength particles reads:
\begin{myequation}\label{scat}
 \kappa_{\mathrm{scat}}(x_p) = \sigma_{\mathrm{scat}} c \frac{f(x_p)}{V_m},
\end{myequation}
with $V_m$ being the mode volume, $c$ the photon velocity, and $\sigma_{\mathrm{scat}} = \frac{k^4}{6 \pi \epsilon_0^2} |\alpha(\omega)|^2$ the Rayleigh scattering cross-section~(where $k = \frac{2 \pi}{\lambda}$ is the wavevector of the incident light).
We begin by comparing the relative effects of the position dependent scattering rate and cavity frequency shift on the intra-cavity photon number. \\ \\
Fig.~2b in the main text shows how the particle motion shifts the resonance peak of the intra-cavity photon number spectrum.
In contrast, a position dependent scattering rate does not shift the peak, but instead alters its width and height.
With this picture in mind we can neglect the effect of the position dependent scattering rate, if the change in scattering rate  $\delta \kappa(x_p)=\kappa_{\mathrm{scat}}(x_p)$ is much smaller than the frequency shift $\delta \omega_c(x_p)$ induced by the same particle movement. Using Eq.~(\ref{scat}) and Eq.~(\ref{cc}) and comparing these two quantities yields:
\begin{myequation}\label{neg}
 \frac{|\delta \kappa_{\mathrm{scat}}(x_p)|}{|\delta \omega_c(x_p)|} \sim (kr)^3 \ll 1,
\end{myequation}
which allows us to neglect the position dependence of the scattering rate for sub-wavelength particles.
\\
\\
Nonetheless we have to consider the reduced quality factor of the resonator-particle system due to  scattering of light. The total cavity decay rate is $\kappa = \frac{\omega_c}{Q} + \kappa_{\mathrm{scat}}(x_0)$ where $Q=\frac{\omega_c}{\kappa_{\mathrm{ex}} + \kappa_{\mathrm{int}}}$ is the quality factor of the empty cavity.
Thus, the back-action parameter reduces to
\begin{myequation}\label{neg2}
\eta =  Q \frac{\alpha(\omega)}{\epsilon_0 V_m} \frac{1}{1 + \frac{ \kappa_{\mathrm{scat}}(x_0)}{\kappa_{\mathrm{ex}} + \kappa_{\mathrm{int}}}}.
\end{myequation}
From Eq.~(\ref{scat}), assuming that $\alpha(\omega) \approx \epsilon_0 V$ and writing $V_m = \nu \left(\frac{\lambda}{2}\right)^3$, where $\nu$ tells us how close the light is focused to the diffraction limit, the scattering rate reads
\begin{myequation}\label{nsc}
\kappa_{\mathrm{scat}}(x_p) \approx \kappa_{\mathrm{scat}}(x_0) \lessapprox \frac{8 \epsilon^2}{27 \pi^2 \nu} (kr)^6 \omega_L.
\end{myequation}
Inserting this into Eq.~(\ref{neg2}) finally yields
\begin{myequation}\label{op}
\eta =   \frac{4}{3 \pi^2} \frac{Q}{\nu} \frac{(kr)^3}{1 + \frac{Q}{\nu} \frac{8}{27\pi^2} (kr)^6},
\end{myequation}
and is plotted in Fig.~\ref{max1} and Fig.~\ref{max2} for $\nu = 1$. In the limit that $\frac{Q}{\nu} \frac{8}{27\pi^2} (kr)^6 \ll 1$, we recover our results from the main text where scattering is negligible, and decreasing the mode volume or increasing the quality factor has the same effect on the back-action parameter. In general, however, for a
given value of $\frac{Q}{\nu}$ for an empty resonator, there is a maximum achievable $\eta$,
\begin{myequation}\label{op2}
\eta_{\mathrm{max}} =   \sqrt{\frac{3 Q}{2 \pi^2 \nu}},
\end{myequation}
which occurs at an optimized particle size of
\begin{myequation}\label{o5}
kr = \sqrt[6]{\frac{27 \pi^2 \nu}{8 Q}}.
\end{myequation}


\section{3 time averaged experienced intensity}

\begin{figure}[h!,t,b]
\centering
\includegraphics[scale=0.71]{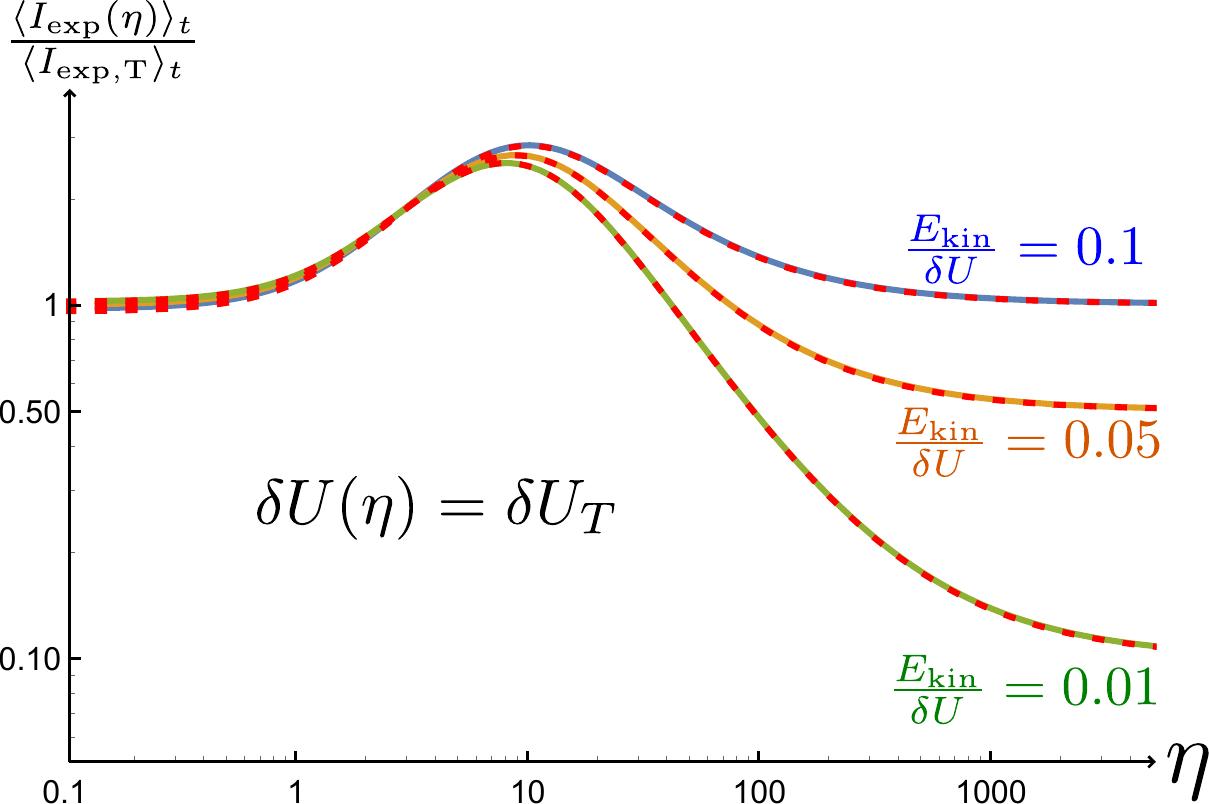}
\caption{Plot of the time-averaged experienced intensity of the particle as a function of back-action parameter $\langle I_{\mathrm{exp}}(\eta) \rangle_t$, normalized by the value in the optical tweezer regime $\eta \ll 1$. The two cases are set to have equal trap depth. The plot is numerically calculated for the case of trapping in the fundamental mode of a Fabry-Perot cavity $f(x) = \cos^2(kx)$ with resonant positions $k x_r= \frac{\pi}{10}$. The red line shows the excellent agreement between Eq.~(\ref{an5}) and the numerical simulation of Eq.~(9) in the main text.}
\label{agree}
\end{figure}

Here we want to derive Eq.~(10) of the main text. In order to do so, we multiply Eq.~(4) with $\frac{f(x)}{f'(x)} \cdot dt$ and integrate both sides over a quarter of an oscillation period:
\begin{myequation}\label{hg}
\int_0^{p_{\mathrm{max}}} dp \frac{f(x)}{|f'(x)|} = \frac{\omega_c \alpha(\omega)}{2 V_m} \int_0^\frac{T}{4}  dt \cdot n(x(t)) \cdot f(x(t))
\end{myequation}
We integrate from the classical turning point (where the momentum is zero) to the trap minimum (where the momentum is maximal) which relates to a quarter of the oscillation period $T$.
Using Eq.~(9) of the main text, the right side of the previous equation is proportional to the time averaged experienced intensity and we can formally rewrite Eq.~\ref{hg} as follows:
\begin{myequation}\label{exa}
\langle I_{\mathrm{exp}} \rangle_t=\frac{4 c}{T \alpha(\omega)} \int_0^{p_{\mathrm{max}}} dp \frac{f(x)}{|f'(x)|}
\end{myequation}
To proceed we make two approximations:
First we approximate the oscillation period $T$ in the high back-action regime as $T \approx 4 \frac{x_t} {v_{\mathrm{max}}}$.
In particular, the particle moves in a square well with length $\delta x = 2 x_r = 2 x_t$, where $x_t$ is the classical turning point, and $v_{\mathrm{max}}$ is the maximum velocity in the middle (minimum) of the potential. Additionally, in the high back-action regime the particle significantly changes its momentum only around the classical turning point $x_t$ when it hits one of the edges of the square well. Since the momentum change occurs in a narrow region, we can approximate in the integral $f(x) \approx f(x_t)$ and $|f'(x)| \approx |f'(x_t)|$.
These approximations lead to the following equation:
\begin{myequation}\label{an}
 \langle I_{\mathrm{exp}} \rangle_t \approx   \frac{c\epsilon_0}{\alpha(\omega)}  \frac{2}{x_t}  \frac{f(x_t)}{|f'(x_t)|} E_{\mathrm{kin}}
\end{myequation}
where $v_{\mathrm{max}} \cdot p_{\mathrm{max}} = 2 E_{\mathrm{kin}}$, with $E_{\mathrm{kin}}$ being the maximal kinetic energy in the trap.
Now we can normalize this time averaged experienced intensity with the time averaged experienced intensity of the optical tweezer regime. We begin with expanding the potential Eq.~(8) for small $\eta$:
\begin{myequation}
U_T(x)= -  2 \hbar \frac{\kappa_{\mathrm{ex}}}{\kappa}  E_0^2  \frac{\eta}{1+\tilde{\Delta}^2} f(x).
\end{myequation}
Since $f(x)$ only varies between 0 and 1, it follows that the trap depth $\delta U_T$ is given by
\begin{myequation}
 \delta U_T =  2 \hbar \frac{\kappa_{\mathrm{ex}}}{\kappa}  E_0^2  \frac{\eta}{1+\tilde{\Delta}^2}.
\end{myequation}
Next we insert Eq.~(7) of the main text, and assume that the particle is tightly trapped ($ k \delta x \ll 1$) around the point of maximum intensity. As the change in intra-cavity photon number is negligible, we can approximate $n(x_p)\approx n(x_0)$. Eq.~(9) from the main text then predicts that
\begin{myequation}\label{ing2}
\langle I_{\mathrm{exp,T}} \rangle_t  =   \frac{c \hbar  \omega_L }{2 V_m }  n(x_0) f(x_0)
\end{myequation}
in the optical tweezer regime. For the case where the particle is trapped around the antinode of the fundamental mode of a Fabry-Perot cavity:
\begin{myequation}\label{an5}
 \langle I_{\mathrm{exp,T}} \rangle_t = \frac{c\epsilon_0}{\alpha(\omega)} \delta U_T.
\end{myequation}
Normalizing Eq.~\ref{an} with Eq.~\ref{an5} and ensuring that $\delta U = \delta U_T$ for all $\eta$ yields:
\begin{myequation}\label{an7}
 \frac{\langle I_{\mathrm{exp}} \rangle_t}{\langle I_{\mathrm{exp,T}} \rangle_t} \approx     \frac{2}{x_t}  \frac{f(x_t)}{|f'(x_t)|} \frac{E_{\mathrm{kin}}}{\delta U}.
\end{myequation}
Surprisingly this equation is valid for all $\eta$ as long as the particle is confined sufficiently close to the antinode. Fig.~S\ref{agree} shows the excellent agreement between the numerical simulation and the analytic solution obtained by Eq.~(\ref{an7}).
Taking the limit $\eta \rightarrow \infty$ of Eq.~(\ref{an}) implies $x_t \rightarrow x_r$ which reproduces Eq.~(10) of the main text:
\begin{myequation}\label{ans}
 \lim_{\eta \rightarrow \infty }  \langle I_{\mathrm{exp}}(\eta) \rangle_t = \frac{2c \epsilon_0}{\alpha(\omega)}\frac{f(x_r)}{|f'(x_r)|} \frac{E_{\mathrm{kin}}}{x_r}
\end{myequation}

\section{4 Optimization of the harmonic back-action regime}
Here we want to maximize the spring constant $k_{\mathrm{opt}}=k_{\mathrm{hb}}+k_{T}$ given by Eq.~(11) in the main text. $k_{\mathrm{hb}}=\sum_i n_i'(x_0) \omega'_{c,i}(x_0)$ describes the first term in Eq.~(11) and originates from changes of photon number with particle position, whereas $k_{T}$ is the familiar term known from optical tweezers.
The optimization is done for a fixed experienced intensity $\langle I_{\mathrm{exp}} \rangle_t$ if we consider two trapping modes of a cavity. As a result we will derive Eq.~(12) and Eq.~(13) in the main text and conclude how to optimally choose the laser detunings for the trapping modes. \\ \\
We focus on the regime where the trap minimum $x_0$ is located roughly at a distance $\sim \frac{1}{k \eta} $ away from both resonant positions, where the photon number $n(x_p)$ can be linearized around the trap minimum $x_0$ for each trapping mode $i$: $n_i(x) \approx n_i(x_0) + n_i'(x_0) (x-x_0)$. A linear change in photon number with displacement implies a harmonic trap, because the force is proportional to the photon number (see Eq.~(4) and Eq.~(6) in the main text and note that $f_i'(x) \approx f_i'(x_0) \approx f_i'(x_{\mathrm{r}i})$ for $k \delta x  \ll 1$).
Using  Eq.~(9) in the main text, the term proportional to $n_i'(x_0)$ does not contribute to the time-averaged intensity due to the harmonic motion. Under these circumstances Eq.~(\ref{ing2}) is valid in the harmonic back-action regime as well and the particle experiences the following time averaged intensity from each trapping mode $i$:
\begin{myequation}\label{insg}
 \langle I_{\mathrm{exp},i} \rangle_t \approx  \frac{2 E_0^2 \kappa_{\mathrm{ex}} c \hbar \omega_L}{ \kappa^2 V_m} \frac{f_i(x_0)}{1+ (\eta f_i'(x_{\mathrm{r}i}))^2(x_0-x_{\mathrm{r}i})^2},
\end{myequation}
where we linearized the mode profiles around their resonant positions in Eq.~(7) in the main text. This is a good approximation if the the width of the intensity peaks is smaller than the spatial variations of the mode profiles, which is the case for $\eta \gg 1$.
Now can write the contributions to the first term $k_{\mathrm{hb}}$ of  Eq.~(11) in the main text as:
\begin{myequation}\label{springkonst}
k_{\mathrm{hb},i} \approx  4 E_0^2 \frac{\kappa_{\mathrm{ex}}}{\kappa} \left(\eta f_i'(x_{\mathrm{r}i}) \right)^2 \frac{\eta |f'_i(x_{\mathrm{r}i})(x_0-x_{\mathrm{r}i})|}{\left(1+ (\eta f_i'(x_{\mathrm{r}i}))^2(x_0-x_{\mathrm{r}i})^2\right)^2}.
\end{myequation}
Expressing the optical tweezer term $k_T$ in the same way, we can write $k_{\mathrm{opt},i}$ in terms of $\langle I_{\mathrm{exp},i} \rangle_t$:

\begin{myequation}\label{kopff}
 k_{\mathrm{opt},i} =  \frac{\alpha(\omega)}{ c \epsilon_0} \langle I_{\mathrm{exp},i} \rangle_t  \frac{1}{f_i(x_0)} \left[ \frac{2r_i}{1+r_i^2} \eta_i f_i'(x_{\mathrm{r}i})^2- f_i''(x_0)\right].
\end{myequation}
$r_i= |\eta_i f_i'(x_{\mathrm{r}i}) (x_{\mathrm{r}i}-x_0)| $ physically describes the ratio between half of the width of an intensity peak $\frac{1}{\eta_i f_i'(x_{\mathrm{r}i})}$ and the distance of the respective resonant position of mode $i$ from the trap minimum $|x_{\mathrm{r}i}-x_0|$.
The spring constant is maximized for $r_i=1$ for which Eq.~(\ref{kopff}) reduces back to Eq.~(12) in the main text.
For $\eta_i \gg 1$ and $r_i= 1$ the contribution to the spring constant proportional to $f_i''(x_0)$ can be neglected and the spring constant purely arises from changes of photon numbers with particle position. In contrast, for $\eta \ll 1$ we can neglect the contribution proportional to $f'_i(x_0)^2$ reaching again the optical tweezer regime. \\
\\
Eq.~(13) in the main text is derived by forming the ratio of these two contributions to the spring constant $\frac{k_{\mathrm{hb}}}{k_T}$ and comparing the two experienced intensities necessary to create the same spring constant in each regime. To derive this, we also assume that the trapping modes consist of the first and second modes of a Fabry-Perot cavity, which have equal back-action parameters $\eta_i$. We also use that $ \langle I_{\mathrm{exp,1}} \rangle_t f_2(x_0) \approx \langle I_{\mathrm{exp,2}} \rangle_t f_1(x_0)$  using Eq.~(4) in the main text with Eq.~(\ref{ing2}) and  $|f_1'(x_0)| \approx |f_2'(x_0)|$ close to the trap minimum.
 \\
 \\
 \section{5 Position dependence of scattering rate in the harmonic back-action regime}

The explicit position dependence of the scattering rate $ \kappa_{\mathrm{scat}}(x_p)$ contributes positively to the optical spring constant in the harmonic back-action regime with an additional term proportional to $f'(x_r)^2$:
\begin{myequation}\label{kopff}
 k_{\mathrm{hb},i} =  \frac{\alpha(\omega)}{ c \epsilon_0} \langle I_{\mathrm{exp},i} \rangle_t  \frac{1}{f_i(x_0)} \left[ \frac{2r_i}{1+r_i^2} \eta_i   + \frac{\frac{2 \delta\kappa_s(x_0)}{\kappa_i f_i(x_0)}}{1+r_i^2}\right] f_i'(x_{\mathrm{r}i})^2
\end{myequation}
where now $\kappa=\kappa_{\mathrm{ex}}+\kappa_{\mathrm{int}}+\kappa_{\mathrm{scat}}(x_0)$
and the back-action parameter is reduced by particle scattering $\eta = \frac{\alpha(\omega)}{\epsilon_0 V_m}\frac{\omega_c}{\kappa}$.
Thus in the harmonic back-action regime we can define an effective back-action parameter
which is the sum of the two terms in the brackets. Interestingly we can choose with laser frequency $r_i=0$, which sets the term coming from frequency shifts to zero. This makes the second term arising from position dependent particle scattering experimentally accessible, which would be a factor $\sim (kr)^3$ smaller than the term arising from frequency shifts in its optimized case ($r_i=1$).

  \bibliographystyle{unsrt}
\bibliographystyle{apsrev}